\documentclass[conference]{IEEEtran}
\IEEEoverridecommandlockouts
\usepackage{cite}
\usepackage{amsmath,amssymb,amsfonts}
\usepackage{algorithmic}
\usepackage{graphicx}
\usepackage{textcomp}
\usepackage{listings}
\usepackage{xcolor}
\def\BibTeX{{\rm B\kern-.05em{\sc i\kern-.025em b}\kern-.08em
    T\kern-.1667em\lower.7ex\hbox{E}\kern-.125emX}}

\begin{document}

\title{Performance Optimization of 3D Stencil Computation on ARM Scalable Vector Extension}

\author{\IEEEauthorblockN{1\textsuperscript{st} Hongguang Chen}
\IEEEauthorblockA{\textit{Department of Computer Science} \\
Chalmers University of Technology\\
Gothenburg, Sweden \\
Email: chenhon@chalmers.se}
% \and
% \IEEEauthorblockN{2\textsuperscript{nd} A2}
% \IEEEauthorblockA{\textit{Department of Computer Science} \\
% The University of Gothenburg\\
% Gothenburg, Sweden \\
% Email: A2@student.gu.se}
% \and
% \IEEEauthorblockN{3\textsuperscript{rd} A3}
% \IEEEauthorblockA{\textit{Department of Computer Science} \\
% Chalmers University of Technology\\
% Gothenburg, Sweden \\
% Email: A3@chalmers.se}
}

\maketitle

\begin{abstract}
Stencil computation is essential in high-performance computing, especially for large-scale tasks like liquid simulation and weather forecasting. Optimizing its performance can reduce both energy consumption and computation time, which is critical in disaster prediction. This paper explores optimization techniques for 7-point 3D stencil computation on ARM’s Scalable Vector Extension (SVE), using the Roofline model and tools like Gem5 and cacti. We evaluate software optimizations such as vectorization and tiling, as well as hardware adjustments in ARM SVE vector lengths and cache configurations. The study also examines performance, power consumption, and chip area trade-offs to identify optimal configurations for ARM-based systems.
\end{abstract}

\begin{IEEEkeywords}
HPC, 3D Stencil, ARM SVE, gem5
\end{IEEEkeywords}

\section{Introduction}
Stencil computation \cite{dubey2014stencils} has been widely researched in high performance field.
It plays a vital role in various tasks, especially large-scale science studies like liquid simulation, and weather forecasting \cite{glinski2001non} \cite{10.1145/3577193.3593719}. Improving its performance reduces energy consumption and saves time, especially in disaster prediction, where prediction time is crucial.

Previous studies \cite{doi:10.1137/070693199} \cite{maruyama2014optimizing} \cite{kamil2005impact} have shown that optimization of stencil computation can be obtained through two primary approaches: hardware and software. On the hardware side, optimizations focus on cache configurations, utilizing heterogeneous hardware, and using specialized computational units. With the rise of GPUs, acceleration on GPUs has also become a mainstream approach. On the software side, significant improvements have been made in code optimization, memory architecture tuning, distributed computing, and multithreading techniques. Additionally, researchers have explored power-performance trade-offs to identify optimal configurations that balance energy efficiency with computational performance.

In \cite{doi:10.1137/070693199}, researchers explored how different memory systems affect 3D stencil computations, such as large on-chip caches, automatic prefetching, and increasing latency. The researchers developed a benchmark to assess prefetching effectiveness in cache-based systems and validated a parameterized probe as a proxy for stencil computations across several processors. They also proposed an analytical memory cost model to predict cache-blocking performance. Findings indicate that recent memory system trends have diminished the effectiveness of traditional cache-blocking optimizations.

However, most of the work in this direction focuses on optimization with GPU \cite{doi:10.1137/070693199} , with a more powerful CPU \cite{maruyama2014optimizing}, like Intel Xeon, and mostly focuses on one dimension, lacking bottom-to-top-level optimization research. 

Arm architecture, due to its’ better balance of power and performance, is becoming the main choice of next-generation computers, computation centers, cloud servers, and even next-generation supercomputers to this day. We have noticed that there are just a few works of optimization from hardware and software in the latest accelerating technique of ARM x64 architecture, Scalable Vector Extension (SVE) \cite{Stephens_2017}, the successor of Neon.

\begin{figure}[htbp]
\centerline{\includegraphics{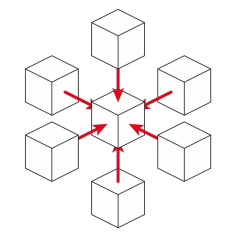}}
\caption{7-point 3D stencil computation}
\label{fig}
\end{figure}

In this paper, we explore the performance relationship between the different hardware configurations, workloads and different code level optimization of 7-point 3D stencil computation, Fig.~\ref{fig}.  This study is based on the Roofline model and employs Gem5 \cite{10.1145/2024716.2024718} and cacti \cite{cacti-tool} to make analyses on different hardware configurations available.

In Section II, we present:

\begin{itemize}
    \item \textbf{Performance Analysis}: Evaluation of 3D stencil computation performance using the Roofline model \cite{10.1145/1498765.1498785}.
    \item \textbf{Software Optimizations}: Techniques such as SVE vectorization, loop unrolling, and tiling to enhance performance.
    \item \textbf{Hardware Optimizations}: Adjustments to ARM SVE vector lengths and cache configurations for improved efficiency.
    \item \textbf{Trade-offs}: Examination of performance versus power consumption and chip area.
\end{itemize}

\section{Method}
\subsection{Simulator and Setting}
Testing the performance of new hardware architectures or configurations is challenging, as it is impractical to fabricate all possible hardware variations for evaluation. Therefore, simulation provides a practical and reliable approach. However, the simulator must be well-designed to ensure that the results accurately reflect real-world performance.

The Gem5 \cite{10.1145/2024716.2024718} is a simulation framework that supports cycle-accurate architectural modeling and provides performance data at varying levels of granularity.

In this study, the simulated system features a 4-core CPU with ARM SVE support, a two-level cache, and DDR3 memory. The theoretical peak performance of each core can be calculated using the following equation:
\begin{equation}
P_{\text{peak}} = f \times \text{fmadd} \times \frac{2048\text{-bit}}{FP32}  = 256 \text{ GFLOPS} \label{eq}
\end{equation}

Where $P_{\text{peak}}$ represents the performance at a clock frequency $f=2 $ GHz, and the \textquoteleft fmadd\textquoteright (float multi-add) operation unit is 2, based on the configuration. The SVE vector length is set to 2048 bits, as specified in [Todo]. Additionally, the peak DDR3 bandwidth is 13 GB/s, according to the gem5 configuration file.

\subsection{Roofline Analysis}
The Roofline model \cite{10.1145/1498765.1498785} is an important way to analyse the bottleneck of a computing system. In this study, we focus on the 7-point 3D stencil algorithm.

\begin{lstlisting}[language=C, caption={Code snippet of 7-point 3D stencil computation}, label={lst:c_code}]
typedef float real_t;
typedef real_t ***arr_t;
void stencil_3d_7point(arr_t A, arr_t B, 
        const int nx,const int ny, const int nz){
  for (int i = 1; i < nx - 1; i++)
    for (int j = 1; j < ny - 1; j++)
      for (int k = 1; k < nz - 1; k++)
        B[i][j][k] = (
          A[i - 1][j][k] + A[i][j - 1][k] +
          A[i][j][k - 1] + A[i][j][k] +
          A[i][j][k + 1] + A[i][j + 1][k] +
          A[i + 1][j][k]) / 7.0;
  // then copy B back to A
}
\end{lstlisting}

In the code snippet Listing~\ref{lst:c_code}, each point involves 7 floating-point operations and 8 memory references (7 reads and 1 write). Under ideal cache conditions (sufficiently large, no cache misses) and optimal vectorization, the cache can effectively filter out all but 1 read and 1 write for each element of both variables A and B. Additionally, the intermediate variable B in the code snippet can be eliminated. Therefore, for a given task size N, under ideal conditions, the ideal arithmetic intensity ($AI_{\text{ideal}}$) can be calculated as follows:

\begin{equation}
\begin{aligned}
AI_{\text{ideal}} &= \frac{7 \text{ op} \times nx \times ny \times nz}{ 2 \text{ ref} \times nx \times ny \times nz \times 4 \text{ Bytes} } \\
                  &= 0.875 \text{ float/Byte}
\end{aligned}
\label{eq}
\end{equation}

The attainable performance (At) of a single core is:
\begin{equation}
\begin{aligned}
At_{\text{ideal}} &= 0.875 \text{ flop/Byte} \times 13 \text{ GB/s} \\ &\approx 11.375 \text{ GFLOPS}
\end{aligned}
\end{equation}

The $At_{\text{ideal}}$ is lower than the peak performance, suggesting a memory performance bottleneck, which can be alleviated by improving memory I/O speed. However, this is not focused in this paper. In practical scenarios, cache performance and optimization are typically inferior to those under ideal conditions. This implies that there is still substantial room for optimization to approach theoretical performance levels.

\subsection{Hardware-level optimization}
This study delves into hardware optimization, focusing on two critical components: the Cache system \cite{kamil2005impact} and the length of the SVE unit.

To identify the effect of the cache system, we assess the performance of various problem sizes while maintaining a fixed cache size. To eliminate the effect of the multi-core mechanism, we use 1 core configuration shown as Table~\ref{tab:p2_gem5_config}.

\begin{table}[htbp]
\centering
\caption{Gem5 Command Parameters}
\label{tab:p2_gem5_config}
\begin{tabular}{|l|l|}
\hline
\textbf{Parameter} & \textbf{Value} \\ \hline
CPU type & \texttt{ex5\_big} \\ \hline
Number of cores & \texttt{1} \\ \hline
L2 cache size & \texttt{64KB} \\ \hline
L1 data cache size & \texttt{8KB} \\ \hline
L1 instruction cache size & \texttt{8KB} \\ \hline
SVE length & \texttt{16 (2048 bits)} \\ \hline
Task sizes & \texttt{[N=5,10,20,40]} \\ \hline
% \multicolumn{4}{l}{$^{\mathrm{a}}$Sample of a Table footnote.}
\end{tabular}
\end{table}

With this configuration, different size workloads are applied, which are smaller than the L1 cache, equal to the L1 cache, larger than the L1 cache and larger than the L2 cache.
The total cache size of the 7-point 3D stencil algorithm can be computed as follows:

\begin{equation}
N \times N \times N \times 4 = 4N^3
\end{equation}

Where 4 is the size of float type. Thus, for an 8KB L1 cache, the maximum feasible $N$ can be computed as:
\begin{equation}
\sqrt[3]{\frac{8KB}{2 \times 4}} \approx 10
\end{equation}

Similarly, for a 64KB L2 cache, the approximate maximum $N$ is 20. The number `2’ signifies the presence of two variables, A and B, which will undergo read and write operations.

To identify the effect of SVE length, we conducted experiments with optimized code with a system configured with various L2 caches from 128 KB to 4096 KB and different SVE lengths from 128 bits to 2048 bits. Additionally, two workloads: N = 32 and N = 64 are tested.

\subsection{Code-level optimization}
When dealing with vector architectures, it's essential to adapt the kernels to leverage their advantages fully.

Several approaches to kernel transformation exist, including auto-vectorization, manual vectorization using intrinsic instructions, and assembly instructions.

For benchmarking purposes, we maintain compilation parameters as \textquoteleft -O3 -fno-tree-vectorize\textquoteright \space
to disable vectorization optimization. To enable auto-vectorization explicitly, we use \textquoteleft-ftree-vectorize\textquoteright. Similarly, for manual vectorization with SVE using intrinsic instructions in Listing\ref{lst:sve_s_7p_3d} (detail code provided in appendix), we use parameters similar to the benchmarks.

\begin{lstlisting}[language=C, caption={SVE instruction}, label={lst:sve_s_7p_3d}]
svcntw()
svwhilelt_b32()
svld1()
svadd_m()
svst1()
\end{lstlisting}

Additionally, when utilizing vectorization, multiple optimizations can be employed, including loop unrolling, loop reordering, loop fusion, or logic changes.

We then execute the stencil code compiled with different parameters on gem5 with 1 core, 1024KB L2 cache, and 64 KB L1 cache with three different task sizes: 16, 32, and 64.

Utilizing multi-core processors with multiple threads is an essential approach to enhance performance. We conducted parallel optimization on the optimal stencil version using OpenMP \cite{dagum1998openmp} and the code is provide in Appendix. we set threads to 1, 4 and 8 and ran the code with a 2048KB L2 cache, 64KB L1 cache, and SVE lengths of 128 bits and 2048 bits, separately. The result is shown in Table \ref{tab:performance_metrics}.

\subsection{Area and Power}
Increasing cache size and vector length can indeed improve performance, but it also raises concerns about the area and power consumption of the chip. Therefore, it is crucial to identify trade-offs that maximize the benefits of longer vector lengths and larger cache sizes while minimizing the associated increase in area and power consumption.

To address this, we refer to the \textquoteleft Area calculations for different vector lengths\textquoteright \space from the appendix, utilizing the Fujitsu A64FX with a 512-bit vector length and 7nm technology as our baseline. The chip area is calculated as follows:

\begin{equation}
\begin{aligned}
\text{Total Chip Area} &= \text{(Rest of Area excluding VPU}\\ &+ \text{Register File)} \\ &+ \text{(VPU + Register File)}
\end{aligned}
\end{equation}

For a single core, the \textquoteleft Rest of Area excluding VPU\textquoteright \space is 1.78 \(mm^2\), and the impact of vector lengths on the VPU + Register File is proportional. Specifically, based on a 512-length VPU with a size of 0.88 \(mm^2\), the size of the VPU with vector lengths ranging from 128 bits to 2048 bits can be computed as follows:

\begin{equation}
Area_x = \frac{Area_{512}}{512} \times 0.88
\end{equation}

According to \cite{9410320}, most of the power consumption is attributed to the cache. Therefore, we conducted tests using the cacti tool \cite{cacti-tool} to evaluate the area and power of different cache sizes.

\section{Result}

\subsection{Performance on different workload}

\begin{figure}[htbp]
    \centering
    \includegraphics[width=0.45\textwidth]{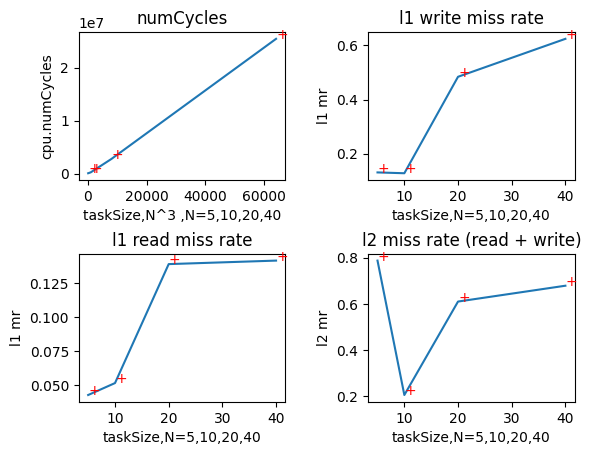}
    \caption{Performance on different workloads with fix cache size}
    \label{fig_p2_1}
\end{figure}
Because cache miss rate is one of important part affecting the computing performance, finding the effect of the cache size associated with workload size will help in selecting the cache sub-system for a specific system.
In the top-left sub-figure of Fig.\ref{fig_p2_1}, we observe a clear relationship between computing time (indicated by \textquoteleft cpu.numCycles\textquoteright) and task size N: as N increases, the computing time rises proportionally. However, the miss rates for L1 and L2 caches exhibit distinct patterns.

For L1 cache, both read and write miss rates remain low when the workload $N < 10$. A sharp increase in these miss rates occurs once the task size surpasses its capacity ($N>10$). In contrast, the overall miss rate for L2 reads and writes displays an unexpected behavior. It reaches its lowest point when $N=10$, while exhibiting significantly higher miss rates both when $N<10$ and $N>10$.

This counterintuitive spike in the L2 miss rate for $N<10$ can be attributed to the limited scale of the workload, where the impact of cold misses becomes prominent \cite{easton1978cold}.

\subsection{Code optimization effect}

\begin{figure}[htbp]
    \centering
    \includegraphics[width=0.45\textwidth]{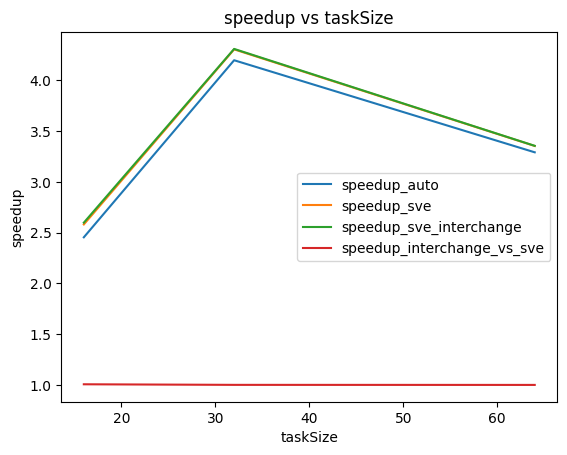}
    \caption{Speed up with different code optimization}
    \label{p3_2}
\end{figure}

\begin{figure}[htbp]
    \centering
    \includegraphics[width=0.45\textwidth]{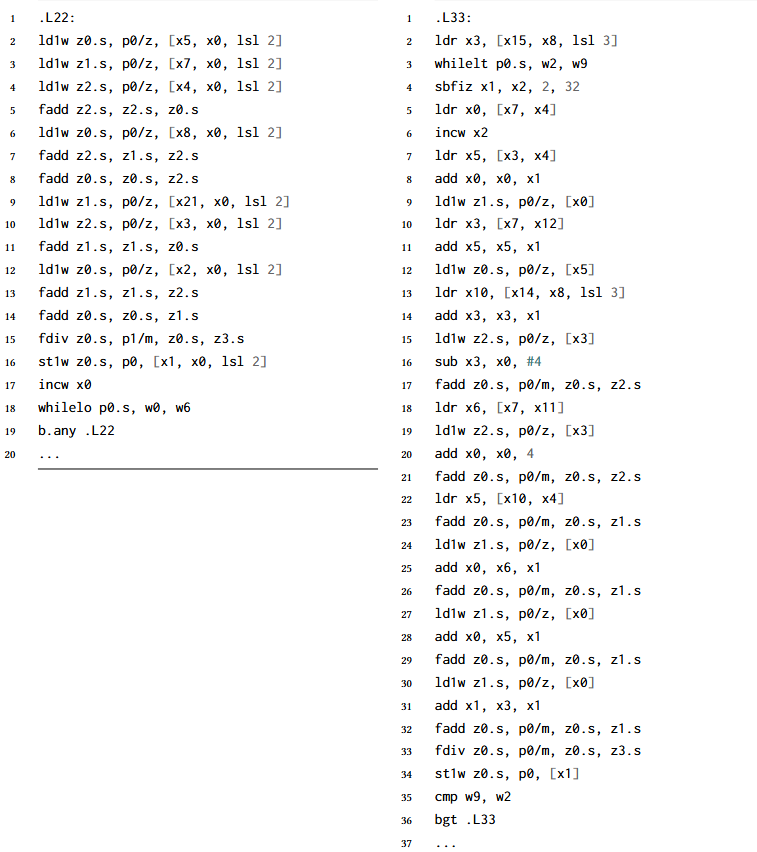}
    \caption{Assembly Code for Different Optimization Parameters: \textquoteleft Auto\textquoteright (Left) and Manual \textquoteleft SVE\textquoteright (Right)}
    \label{code_compare}
\end{figure}
Optimization from code is an effective and efficiency approach.
Fig.\ref{p3_2} displays the speedup relative to 3 types of code-level optimization with three different task sizes N: 16, 32, and 64. Results indicate that auto-vectorization and manually optimized versions outperform the benchmark (without any optimization), achieving speedups ranging from 2.5 to 4.5 times across different task sizes. Notably, the SVE version exhibits only slightly better performance than the auto-vectorized version. This observation could be explained by the relatively simple nature of the kernel for the 7-point stencil, which involves summing seven independent points. Additionally, Fig.\ref{code_compare} shows the code generated by the compiler for the ‘auto’ and the ‘SVE’ version. We can see that the compiler can generate highly optimized code.

% \begin{figure}[htbp]
%     \centering
%     \includegraphics[width=0.45\textwidth]{Imgs/p3_3.png}
%     \caption{7-point 3D stencil computation}
%     \label{fig}
% \end{figure}

\subsection{Hardware effect}

\begin{figure}[htbp]
    \centering
    \includegraphics[width=0.45\textwidth]{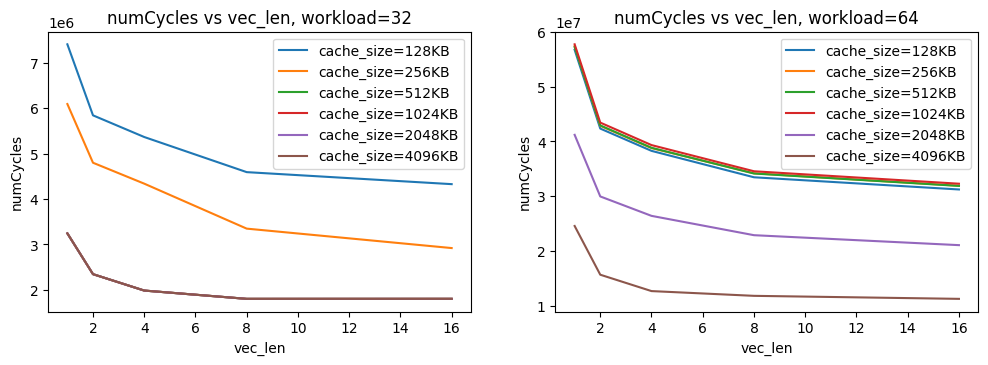}
    \caption{Cache Size, Vector Length, and Performance}
    \label{p4_1}
\end{figure}

\begin{table}[htbp]
    \centering
    \caption{Performance Metrics for Different Vector Lengths and Thread Counts}
    \begin{tabular}{|c|c|c|c|}
        \hline
        \textbf{vec\_length} & \textbf{threads} & \textbf{cpu0.numCycles} & \textbf{Speedup (vs thread=1)} \\
        \hline
        128 & 1 & 41,187,344 & - \\
        128 & 4 & 23,662,492 & 1.74 \\
        128 & 8 & 18,133,950 & 2.27 \\
        \hline
        2048 & 1 & 21,048,412 & - \\
        2048 & 4 & 11,554,210 & 1.82 \\
        2048 & 8 & 10,267,740 & 2.05 \\
        \hline
    \end{tabular}
    \label{tab:performance_metrics}
\end{table}
SVE as the main accelerating feature being studied in this paper, the effect of it with different cache setting and thread number is tested. 
Fig.\ref{p4_1} shows the performance of the system with SVE length from 1 to 16, and L2 cache size from 128KB to 4096KB on workload N=32 and N=64, separately.
We can notice that increasing the SVE length can improve the performance, especially when the SVE length is small( From 1 to 8 in the x-axis in  Fig.~\ref{p4_1}). Cache size has a similar effect on the performance.
For a fixed cache size, performance generally improves as the SVE length increases, particularly when the vector length is smaller than 512 bits (4 in the x-axis in Fig.\ref{p4_1}), although at a decreasing rate as the SVE length increases further.

Table \ref{tab:performance_metrics} summarizes the performance results across different combinations of SVE lengths (128-bit, 2048-bit) and thread counts (1, 4, 8). The right column presents the speedup relative to the baseline configuration of a single thread for both SVE lengths.The results indicate that increasing the thread count at a fixed SVE length improves performance. However, the rate of improvement diminishes as the thread count rises. Interestingly, with a fixed number of threads, cases with smaller SVE lengths show relatively higher performance gains.

This phenomenon can be explained by Amdahl’s Law, which is expressed as:

\begin{equation}
\text{speedup} = \frac{1}{f + \frac{1-f}{N}}
\end{equation}

where f represents the fraction of the program that is serial, and N is the number of processors. As N increases, the rate of speedup improvement slows down, approaching a theoretical maximum limit.

\subsection{Chip size and power}
\begin{figure}[htbp]
    \centering
    \includegraphics[width=0.45\textwidth]{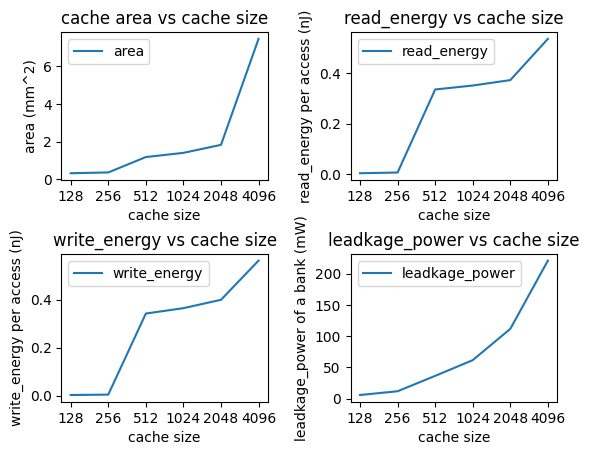}
    \caption{Cache power consumption}
    \label{p4_3}
\end{figure}

The trade-off between performance and energy consumption is another part this paper focuses on. 
Fig.\ref{p4_3} illustrates the area, power, and leakage of the L2 cache for various cache sizes.
It is notable that the area increases rapidly and disproportionately when the size exceeds 2048KB. Moreover, both read and write energy nearly double when the cache size surpasses 256KB. Furthermore, the rate of increase in leakage power significantly accelerates as the cache size increases. Overall, increasing the cache size can improve performance. However, beyond a certain threshold, the additional area and power consumption outweigh the performance gains, diminishing the cost-effectiveness of further cache enlargement.

\section{Conclusion and Discussion}
Our research proved detailed research on the optimization of 3D stencil computing on ARM SVE Architecture, from hardware to software level, addressing a gap that few studies paid attention to in this field, which is the foundation of high-performance science computing, under the condition that most of the researcher focus on optimizing it for GPUs and high-end x86 CPUs.

In the present study, our results show that both hardware configurations (cache sizes, SVE lengths) and code-level strategies (vectorization, loop unrolling, interchanging, multi-thread parallelizing) impact computational efficiency.  Specifically, findings indicated that the SVE lengths contribute more to performance gains when aligned with workload size while miss rates up to a threshold. 

Comparing our findings with previous research, our study confirms the importance of cache configurations and SVE length while expanding on them by providing a more detailed view within the ARM SVE context. Unlike studies focusing solely on cache-blocking \cite{kamil2005impact} or GPU-specific optimizations \cite{doi:10.1137/070693199}\cite{maruyama2014optimizing}, this research demonstrated a bottom-top approach for ARM SVE, integrating hardware and software optimizations.

The limitations of this study include the inability to perform tests on actual ARM SVE hardware due to resource constraints, which may slightly impact the accuracy of the simulation in reflecting real-world conditions. Furthermore, the results here are specific to the 7-point stencil computation, whereas real-world applications often involve more complex workloads that demand larger memory resources. Again, distributed systems can influence performance significantly and were not considered in this study. Lastly, although multi-core systems are standard in high-performance computing, we did not investigate the effects of multi-core configurations, which is another area for future investigation.

In conclusion, this study contributes to optimizing 3D stencil computations on ARM SVE by identifying  hardware and software configurations that maximize performance within power and area constraints. These findings advance the field’s understanding of ARM SVE capabilities and lay the groundwork for further exploration, particularly relevant as ARM gains traction in HPC applications.

\section*{Acknowledgment}
I completed the writing of this paper while at Tokyo University, with valuable feedback on the draft from Dr. Yuka Akiyama and Anubhav.

\vspace{12pt}
% \color{red}
% (will remove when everything is done)
% IEEE conference templates contain guidance text for composing and formatting conference papers. Please ensure that all template text is removed from your conference paper prior to submission to the conference. Failure to remove the template text from your paper may result in your paper not being published.
\end{document}